\documentstyle[sprocl,epsf,epsfig,wrapfig]{article}

\def\Journal#1#2#3#4{{#1} {\bf #2}, #3 (#4)}

\def\ZPC{{\em Z. Phys.} C}
\def\be{\begin{equation}}
\def\ee{\end{equation}}
\def\bea{\begin{eqnarray}}
\def\eea{\end{eqnarray}}
\begin{document}
\enlargethispage{10.0cm}

\begin{flushright}
\vspace{-3.0cm}
{\bf UCL/HEP 97-03}
\end{flushright}
\vspace{.8cm}
\title{ENERGY FLOWS AND JET PRODUCTION IN TAGGED $e\gamma$ EVENTS AT LEP
}

\author{ A.M. ROOKE FOR THE OPAL COLLABORATION }

\address{Department of Physics, UCL, Gower Street,
London WC1E 6BT, UK.}


\maketitle
 \vspace{-0.5cm}
\begin{wraptable}[5]{r}{5.5cm}
\vspace{-.55cm}
\begin{tabular}{|l|l|l|l|} \hline
 & 0 jet & 1 jet & 2 jet \\ \hline
{$\!$\rm Data} & 30.7$\%$ & 63.8$\%$ & 5.4$\%$ \\
{$\!$\sc Herwig$\!$} & 34.0$\%$ & 63.6$\%$ & 2.4$\%$ \\
{$\!$\sc Pythia$\!$} & 32.8$\%$ & 65.5$\%$ & 1.7$\%$ \\ \hline
\end{tabular}
\vspace{-0.35cm}
\caption{Jet rates of {\sc Herwig} and {\sc Pythia}
compared to the data.}
\label{tab:proc_01}
 \end{wraptable}
\vspace{-0.1cm}
\section{Introduction}
\vspace{-0.3cm}
It has been shown~\cite{opaleflow} that the predictions of 
the hadronic energy flow in $e\gamma$ DIS processes by  
the QCD-based Monte Carlo generators, 
{\sc Herwig} and {\sc Pythia}, disagree with the data in the
$Q^2$ = 6--30 ${\rm GeV^{2}}$ region. The disagreements are particularly
marked in the regions of pseudorapidity well measured by the OPAL
detector and hence where jets can be accurately reconstructed. 
\vspace{-0.4cm}
\section{Results And Conclusions}
\vspace{-0.15cm}
\begin{wrapfigure}[14]{r}{6.4cm}
\vspace{-0.45cm}
\hspace{0.3cm}
 \epsfig{file=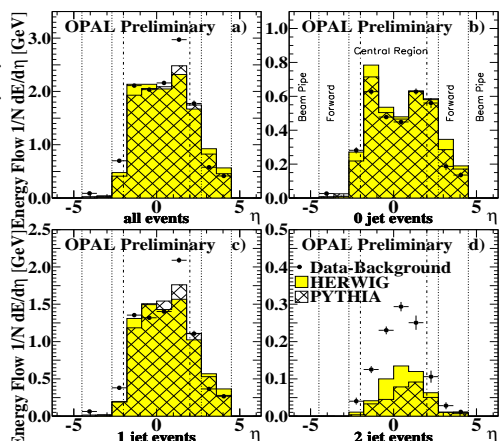,height=5.7cm,width=6.8cm,
bbllx=30,bblly=0,bburx=550,bbury=480}
 \vspace{-1.1cm}
 \caption{Average energy flow in an event as a function of
pseudorapidity $\eta=-\ln{\tan{\frac{\theta}{2}}}$}
 \label{fig:proc_01}
 \end{wrapfigure}
\vspace{-0.1cm}
Table~\ref{tab:proc_01} shows that the fraction of data events with
2 cone jets (of transverse energy ${\rm E_{\rm T,jet} > 3 GeV}$ 
and pseudorapidity ${\rm |\eta_{\rm jet}| < 2}$) 
   exceeds 
 predictions from the {\sc Herwig} and {\sc Pythia}
samples by over a factor of 2.
Fig.~\ref{fig:proc_01}a) shows the average hadronic energy flow
per event as a function of pseudorapidity  and
b), c)  and d) show the contributions to a) from events 
with 0, 1 or 2 jets. The underestimation by {\sc Herwig} and {\sc Pythia}
of events with 2 jets combined with their imperfect modelling of energy flow
(particularly in the pseudorapidity region $0.9<\eta<1.8$ for events with 1 jet) 
give rise to the previously reported discrepancies~\cite{opaleflow} shown in
fig.~\ref{fig:proc_01}a).  
By selecting events with 2 identified hadron cone jets, we have found a subset 
of the data in which disagreements with Monte Carlo predictions are
greatly enhanced. This result has been used to constrain improvements to the
models~\cite{jan}.

\end{document}